\documentclass[twocolumn,preprintnumbers,amsmath,amssymb]{revtex4}

\usepackage{graphicx}
\usepackage{dcolumn}
\usepackage{bm}

\begin{document}

\preprint{to appear in Appl. Phys. Lett.}

\title{
Enhanced thermoelectric properties of Na$_x$CoO$_2$ whisker crystals 
}

\author{G.~Peleckis}

\author{%
T.~Motohashi, M.~Karppinen 
}

\author{H.~Yamauchi}
 \thanks{%
Corresponding author: Prof. Hisao Yamauchi \\
Materials and Structures Laboratory, Tokyo Institute of Technology \\
4259 Nagatsuta, Midori-ku, Yokohama 226-8503, Japan \\
Phone: +81-45-924-5315, Fax: +81-45-924-5365 \\
Electronic mail: yamauchi@msl.titech.ac.jp
}

\affiliation{%
Materials and Structures Laboratory, 
Tokyo Institute of Technology, Yokohama 226-8503, Japan \\
}




\date{%
\today
}

\begin{abstract}
Single-crystalline whiskers of thermoelectric cobalt oxide, 
Na$_x$CoO$_2$, have been grown by an unconventional method from 
potassium-containing compositions, and their transport properties, 
and chemical compositions were determined. 
The growth mechanism was analyzed and interpreted by means of 
thermogravimetric analysis. At 300 K, electrical resistivity $\rho$ 
and thermoelectric power $S$ of the whisker are 102 $\mu\Omega$ cm 
and 83 $\mu$V/K, respectively. 
The power-factor, $S^2/\rho$, thus is enhanced up to $\sim$ 
68 $\mu$W/K$^2$ cm. 
\end{abstract}

\maketitle

Thermoelectric materials that can convert heat flow directly to 
electric flow/current and vice versa are considered to play a key role 
in ``clean'' energy utility of the next generation \cite{Mahan97a}. 
However, requirements for high-efficiency thermoelectric materials are 
not easily satisfied. 
The thermoelectric efficiency is evaluated by the following 
figure-of-merit: $Z = S^2 / \rho \kappa$, where $S$, $\rho$, 
and $\kappa$ denote thermoelectric power, electrical resistivity, 
and thermal conductivity, respectively. 
Therefore, large $S$ values and small $\rho$ and $\kappa$ values 
are simultaneously required to realize high thermoelectric efficiency. 
For the last few years thermoelectric oxides have been widely 
investigated owing to their high chemical stability at temperatures 
higher than $\sim$800 K. 
The sodium-cobalt oxide, Na$_x$CoO$_2$, is one of the most promising 
candidates among the thermoelectric oxides. 
Terasaki {\it et al.} \cite{Terasaki97a} found that this compound 
(originally expressed as NaCo$_2$O$_4$) exhibits simultaneously 
high thermoelectric power and low resistivity 
($S$ $\approx$ 100 $\mu$V/K and $\rho$ $\approx$ 
200 $\mu\Omega$ cm at 300 K). 
However, due to its two-dimensional crystal structure, the compound 
has large electromagnetic anisotropy \cite{Terasaki97a}, 
and thus polycrystalline samples always show higher resistivity values 
($\rho$ $\approx$ 3 m$\Omega$ cm at 300 K \cite{Terasaki00a}). 
For reducing the resistivity and thereby enhancing the $Z$ value of 
Na$_x$CoO$_2$, it is indispensable to develop a method to prepare textured 
or single-crystalline materials. 

Whisker-type single crystals are of significant importance 
due to the following reasons: 
(1) the one-dimensional shape of whiskers should be favorable for 
some synthetic textures, and (2) whiskers may be convenient for 
some tiny-dimensional applications such as natural ``thermoelectric 
wires'' taking advantage of their characteristic shape. 
It was shown that whiskers of Bi$_2$Sr$_2$Co$_2$O$_y$ 
\cite{Funahashi01a,Funahashi02a} exhibit significantly 
enhanced thermoelectric properties as compared with polycrystalline samples. 
Moreover, whiskers are known to be high-quality single crystals with 
low defect densities. 
Therefore, precise investigations on the crystal structure and 
physical properties of Na$_x$CoO$_2$ can be made using whisker crystals. 
In this letter we report thermoelectric properties of Na$_x$CoO$_2$ 
whiskers prepared using an original growth technique. 

It is known that Na$_x$CoO$_2$ forms plate-like crystals 
when NaCl is used as a flux. 
To obtain Na$_x$CoO$_2$ whiskers we used potassium-containing compositions. 
Dry powders of Na$_2$CO$_3$, K$_2$CO$_3$, and Co$_3$O$_4$ were mixed 
with cation ratios, Na : K : Co = 0.525 : 0.225 : 1.00, and calcined 
at 750$^\circ$C in air employing a ``rapid heat-up'' technique 
\cite{Motohashi01a} to avoid evaporation of the alkali metals. 
The calcined powder was ground, pressed into pellets, and fired at 
900$^\circ$C for 12 h in air. 
During the firing, partial melting apparently occurred, and whiskers 
were found to form at the surface of the pellets (Fig.~\ref{SEM}).

Figure~\ref{XRD} shows an x-ray powder diffraction pattern 
(measured with Cu$K_{\alpha1}$ radiation) for pulverized whiskers. 
All the peaks in the pattern can be indexed according to the 
$\gamma$-Na$_x$CoO$_2$ phase \cite{Fouassier73a}. 
Crystal structure analysis using a four-circle diffractometer 
and synchrotron radiation revealed that the whiskers possess 
a hexagonal structure with space group, $P6_3/mmc$ (No.~194) 
\cite{Hanashima02a}, 
being in agreement with the neutron diffraction data for 
a polycrystalline sample of Na$_{0.74}$CoO$_2$ \cite{Balsys96a}. 
The grown whiskers are up to 1.6 mm in length, 15 - 40 $\mu$m 
in width, and 1.5 - 4.0 $\mu$m in thickness: 
the shape is strongly one-dimensional in noticeable contrast to 
plate-like crystals grown by a NaCl-flux method 
\cite{Terasaki97a,Fujita01a}. 
The growth direction of the whiskers was found to be along the ab plane, 
while the smallest dimension is along the $c$ axis. 

Electrical resistivity ($\rho$) and thermoelectric power ($S$) 
of the whiskers were measured: 
$\rho$ was measured using a four-point-probe apparatus 
(Quantum Design; PPMS) with a dc current of 5 mA along the growth 
(i.e. in-plane) direction, and $S$ using a steady-state 
technique with a typical temperature gradient of 0.5 K/cm. 
Figure~ref{r-and-S}(a) shows the dependence of $\rho$ on 
temperature for a Na$_x$CoO$_2$ whisker. 
The $\rho$ value decreases with decreasing temperature, 
exhibiting typical metallic behavior (i.e. $d\rho/dT > 0$) 
in the whole temperature range. 
This is in accordance with the results reported for single crystals 
in previous works \cite{Terasaki97a,Fujita01a}. 
An anomaly in resistivity is observed at $T_t$ $\approx$ 290 K, 
which is considered to be related to a first-order transition 
\cite{Tojo02a}, 
while such an anomaly has not been observed for single-crystalline 
samples of Na$_x$CoO$_2$ \cite{Terasaki97a,Fujita01a}. 
The absolute value of resistivity for the present whiskers 
is by a factor of two smaller than that for single crystals 
of Na$_x$CoO$_2$ grown with the NaCl flux method 
(cf. $\rho$ $\approx$ 200 $\mu\Omega$ cm at 300 K) \cite{Terasaki97a}. 
Figure~\ref{r-and-S}(b) shows the dependence of $S$ on temperature 
for a Na$_x$CoO$_2$ whisker. 
Both the absolute value and the dependence on temperature of $S$ 
are in good agreement with those previously reported for 
single-crystalline and polycrystalline samples 
\cite{Terasaki97a,Terasaki00a}. 
The values of $\rho$ and $S$ at 300 K for the present whisker 
crystal are 102 $\mu\Omega$ cm and 83 $\mu$V/K, respectively. 
The power-factor ($S^2/\rho$) is then calculated at 
$\sim$ 68 $\mu$W/K$^2$ cm, being as large as that for a crystal 
grown by the NaCl-flux method \cite{Terasaki97a} and by a factor of 
$\sim$ 30 larger than that for polycrystalline samples \cite{Terasaki00a}. 

In order to determine the chemical composition of grown whiskers, 
we performed ICP-AES (Seiko Instruments: SPS-1500VR) 
and SEM-EDS (Hitachi: S-4500) analyses. 
In Table~\ref{composition}, the actual cation composition of the whiskers 
as determined by the two independent analysis methods is given 
together with those of the calcined powder and the fired pellets. 
Neither the whiskers nor the pellets contained potassium, 
whereas the actual composition of the calcined powder 
was nearly the same as the nominal one. 
This result indicates that fast evaporation of potassium occurs 
during firing (at $\sim$ 900$^\circ$C), which is apparently essential 
for the growth of whiskers. 
The chemical formula determined for the whisker (in terms of cation 
composition) is Na$_{0.5}$CoO$_2$, being highly sodium deficient 
and free from potassium. 

To investigate the whisker formation mechanism, thermogravimetric 
(TG) measurements were performed. 
A powder mixture of raw materials of 4.295 mg was heated with a rate 
of 3.2$^\circ$C/min in a thermobalance (Perkin Elmer: Pyris 1) 
with an isothermal heating period of 12 h at 900$^\circ$C. 
The TG curve is shown in Fig.~\ref{TG}. Since K$_2$CO$_3$ is highly 
hygroscopic, the first weight loss about 100$^\circ$C is attributed 
to the evaporation of water. 
The second weight loss is seen starting about 500$^\circ$C. 
This is most likely due to the decarbonation process, though the total 
weight loss in this region ($\sim$ 6.4 \%) is a little too large to be 
explained with decarbonation (5.8 \%) only. 
Therefore it is considered that a small amount of alkali metal(s) 
evaporates as well. 
The two weight drops about 100$^\circ$C and 500$^\circ$C were also seen 
for the potassium-free sample of Na$_{0.55}$CoO$_2$ \cite{Motohashi01a}. 
After the large weight drop due to decarbonation, a temperature window 
was found to appear about the calcination temperature (650-830$^\circ$C) 
where the TG curve is relatively flat. 
This should correspond to the stability region of the 
Na$_x$K$_y$CoO$_2$ phase. 
In fact, x-ray diffraction measurements showed that the powder product 
after calcination at 750$^\circ$C is of single phase of hexagonal 
Na$_x$K$_y$CoO$_2$ phase, whereas chemical analyses for the cation 
composition revealed essentially no deviation from the nominal 
potassium content (see Table~\ref{composition}). 
Upon increasing the temperature up to 900$^\circ$C an additional 
weight loss is seen indicating that the fast evaporation of potassium 
from the Na$_x$K$_y$CoO$_2$ phase starts to occur 
in this temperature range. 
In the inset of Fig.~\ref{TG} the normalized weight 
is plotted against time. 
During the isothermal heating period of 12 hours at 900$^\circ$C 
the weight loss continues until potassium completely gets depleted 
from the sample: 
the evaporation rate of potassium from Na$_x$K$_y$CoO$_2$ was estimated 
at $\approx$ 6.5 mol\%/h. 
The TG result is in good agreement with the ICP analysis data 
(Table~\ref{composition}), 
according to which no potassium was found to remain in both the pellets 
and the whiskers. 
Taking into account the results of TG and ICP-AES analyses, 
it is most likely that a large amount of potassium evaporation 
induces partial melting at the surface of the pellet, 
and the partially-melted surface plays an essential role 
in the whisker growth. 

In summary, we have developed an unconventional method to grow whisker 
single crystals of Na$_x$CoO$_2$. 
Using potassium-containing compositions, whisker crystals of 
Na$_x$CoO$_2$ are found to form after firing at 900$^\circ$C. 
At 300 K, $\rho$ and $S$ of the whiskers are 102 $\mu\Omega$ cm 
and 83 $\mu$V/K, respectively. 
Calculated power factor ($S^2/\rho$) at this temperature 
shows a record-high value of $\sim$ 68 $\mu$W/K$^2$ cm 
for the sodium-cobalt-oxide system. 
From ICP-AES and SEM-EDS analyses, it has been revealed that 
the grown whiskers are highly sodium deficient ($x$ $\approx$ 0.50) 
and free of potassium. 
Increasing sodium content up to 0.75 can probably lead to even higher 
thermoelectric power ($S^2/\rho$) values, therefore preparation of 
whiskers with higher sodium contents are on the way in our laboratory.

We are grateful to Prof. I.~Terasaki of Waseda University for 
his contribution in thermoelectric power measurements. 
The present work was supported by Grants-in-aid for Scientific Research 
(Contract Nos. 11305002 and 14038219) from the Ministry of Education, 
Culture, Sports, Science and Technology of Japan.

\newpage

\begin{table}[htbp]
\begin{center}
\caption{%
Chemical compositions of calcined powder, pellets after firing 
at 900$^\circ$C, and grown whiskers as determined 
by ICP-AES and SEM-EDS analyses. 
The given values are normalized by cobalt content. \\
}
\label{composition}

\begin{tabular}{ccccccc}
\hline\hline
Sample & \multicolumn{2}{c}{Nominal} & \multicolumn{2}{c}{ICP-AES}
& \multicolumn{2}{c}{SEM-EDS} \\
\hline\hline
Calcined powder & Na & 0.525 & Na & 0.540 & \multicolumn{2}{c}{---} \\
                & K  & 0.225 & K  & 0.240    &    &          \\
\hline
Pellet          & Na & 0.525 & Na & 0.547    & Na & 0.532    \\
                & K  & 0.225 & K  & $<$0.001 & K  & $<$0.001 \\
\hline
Whisker         & Na & 0.525 & Na & 0.523    & Na & 0.487    \\
                & K  & 0.225 & K  & $<$0.001 & K  & $<$0.001 \\
\hline\hline
\end{tabular}
\end{center}
\end{table}

\newpage

\begin{figure}
\caption{%
Scanning electron micrograph of the grown Na$_x$CoO$_2$ whiskers. 
}
\label{SEM}

\caption{%
X-ray powder diffraction pattern for pulverized 
whisker crystals of Na$_x$CoO$_2$. 
}
\label{XRD}

\caption{%
(a) Temperature dependence of resistivity ($\rho$) 
for a Na$_x$CoO$_2$ whisker. 
An anomaly is seen at $T_t$ $\approx$ 290 K. 
(b) Temperature dependence of thermoelectric power ($S$) 
for a Na$_x$CoO$_2$ whisker.
}
\label{r-and-S}

\caption{%
Thermogravimetric curve of a raw material powder mixture 
(with cation ratios, Na : K : Co = 0.525 : 0.225 : 1.00) 
upon heating in flowing O$_2$ gas. 
(1 - Evaporation of H$_2$O; 2 - Decarbonation process; 
3 - Fast K evaporation) 
The inset represents the same thermogravimetric data 
with respect to time. 
}
\label{TG}

\end{figure}


\begin{thebibliography}{00}

\bibitem{Mahan97a}
G. Mahan, B. Sales, and J. Sharp, 
Phys. Today {\bf 50}, 42 (1997). 

\bibitem{Terasaki97a}
I. Terasaki, Y. Sasago, and K. Uchinokura, 
Phys. Rev. B {\bf 56}, R12 685 (1997). 

\bibitem{Terasaki00a}
I. Terasaki, 
in {\it Proceedings of the 18th International Conference 
on Thermoelectrics (ICTf99)}, Baltimore, MD, USA 1999 
(IEEE, Piscataway, 2000), pp. 569-576. 

\bibitem{Funahashi01a}
R. Funahashi, and I. Matsubara, 
Appl. Phys. Lett. {\bf 79} 362 (2001).

\bibitem{Funahashi02a}
R. Funahashi, and M. Shikano, 
Appl. Phys. Lett. {\bf 81} 1459 (2002).

\bibitem{Motohashi01a}
T. Motohashi, E. Naujalis, R. Ueda, K. Isawa, 
M. Karppinen, and H. Yamauchi, 
Appl. Phys. Lett. {\bf 79}, 1480 (2001). 

\bibitem{Fouassier73a}
C. Fouassier, G. Matejka, J.-M. Reau, and P. Hagenmuller, 
J. Solid State Chem. {\bf 6}, 532 (1973). 

\bibitem{Hanashima02a}
T. Hanashima, G. Peleckis, T. Motohashi, S. Sasaki, 
M. Karppinen, and H. Yamauchi, 
unpublished work at Tokyo Institute of Technology (2002). 

\bibitem{Balsys96a}
R.J. Balsys and R.L. Davis, 
Solid State Ionics {\bf 93}, 279 (1996). 

\bibitem{Fujita01a}
K. Fujita, T. Mochida, and K. Nakamura, 
Jpn. J. Appl. Phys. {\bf 40}, 4644 (2001). 

\bibitem{Tojo02a}
T. Tojo, H. Kawaji, T. Atake, Y. Yamamura, M. Hashida, and T. Tsuji, 
Phys. Rev. B {\bf 65}, 052105 (2002). 

\end{thebibliography}
\end{document}